\documentclass[twocolumn,showpacs,preprintnumbers,amsmath,amssymb,superscriptaddress,prl]{revtex4}

\usepackage{graphicx}
\usepackage{natbib}

\begin{document}

\title{A Quantized $\nu =5/2$ State in a Two-Subband Quantum Hall System}
\author{J.~Nuebler}
\affiliation{Max-Planck-Institute for Solid State Research, Heisenbergstr. 1, D-70569 Stuttgart, Germany}

\author{B. Friess}
\affiliation{Max-Planck-Institute for Solid State Research, Heisenbergstr. 1, D-70569 Stuttgart, Germany}

\author{V.~Umansky}
\affiliation{Braun Centre for Semiconductor Research, Department of Condensed Matter Physics, Weizmann Institute of Science, Rehovot 76100, Israel}

\author{B. Rosenow}
\affiliation{Institute for Theoretical Physics, University of Leipzig, Vor dem Hospitaltore 1, D-04103 Leipzig, Germany}

\author{M. Heiblum}
\affiliation{Braun Centre for Semiconductor Research, Department of Condensed Matter Physics, Weizmann Institute of Science, Rehovot 76100, Israel}

\author{K. von Klitzing}
\affiliation{Max-Planck-Institute for Solid State Research, Heisenbergstr. 1, D-70569 Stuttgart, Germany}

\author{J. Smet}
\email{j.smet@fkf.mpg.de}
\affiliation{Max-Planck-Institute for Solid State Research, Heisenbergstr. 1, D-70569 Stuttgart, Germany}

\date{\today}

\begin{abstract}
The evolution of the fractional quantum Hall state at filling 5/2 is studied in density tunable two-dimensional electron systems formed in wide wells in which it is possible to induce a transition from single to two subband occupancy. In 80 and 60 nm wells, the quantum Hall state at 5/2 filling of the lowest subband is observed even when the second subband is occupied. In a 50 nm well the 5/2 state vanishes upon second subband population. We attribute this distinct behavior to the width dependence of the capacitive energy for intersubband charge transfer and of the overlap of the subband probability densities.
\end{abstract}

\pacs{73.43.-f, 73.63.Hs, 73.43.Qt}

\maketitle

{\em Introduction and key findings - }
Currently, there is strong interest in the even-denominator fractional quantum Hall state (FQHS)~\cite{PhysRevLett.59.1776, PhysRevLett.83.3530} at filling factor $\nu = 5/2$, partially because of its potential relevance for topological quantum computation resulting from the non-Abelian statistics its quasi-particle excitations are predicted to obey~\cite{PhysRevLett.94.166802}. The 5/2 state is usually studied in GaAs/AlGaAs-heterostructures with a single heterointerface or relatively narrow quantum wells (QWs) where electrons occupy only the first subband (1SB). By widening the quantum well the physics is enriched. The accessible density range in these samples is large enough to substantially populate the second subband (2SB). This adds an additional degree of freedom and produces novel ground states as exemplified by the appearance of quantum Hall states at filling 1/2~\cite{PhysRevLett.68.1379} and 1/4~\cite{PhysRevLett.101.266804} as well as quantum Hall ferromagnetic phases absent in single subband systems~\cite{Piazza_QHferromagnet, PhysRevLett.87.196801, PhysRevLett.86.2412}. The total filling factor $\nu_T$, i.e. the ratio of the electron density $n$ and the Landau level (LL) degeneracy, is then the sum of the subband fillings:  $\nu_T = \nu_{\rm 1SB} + \nu_{\rm 2SB}$.  Recently, the 5/2 state was investigated in such wide QWs ~\cite{PhysRevLett.105.246805, liu_thirds, liu_robustness_52}. The incompressible 5/2 quantum Hall state was observed when the 2SB was empty, but was lost upon populating it. Instead a compressible composite fermion liquid formed in the half filled lowest LL of the 2SB while $\nu_{\rm 1SB} = 2$. This literature also suggested that the 5/2 state is stabilized just before the 2SB is occupied~\cite{liu_robustness_52}.

Here, we report that in very wide QWs, the 2D system formed by the 1SB condenses in the 5/2 fractional quantum Hall state even when the 2SB {\em is  occupied}. The 5/2 state continues to exist as an incompressible fluid over a wide range of fillings of the 2SB. This is different from previous reports on two-SB systems~\cite{liu_thirds, PhysRevLett.87.196801} where quantum Hall features follow the {\em total} filling factor. We attribute the dramatic difference between previous reports and our data to the increased QW width in our samples resulting in a larger spatial separation of the charge distributions of the two SBs. This suppresses intersubband scattering and increases the Coulomb energy involved when transferring charge between the two SBs. The existence of a quantized 5/2 state in close proximity to an independent 2DES of variable density (formed by the 2SB electrons) may enable to study their interaction.

{\em Samples and measurements - } Magnetotransport experiments were carried out on $400\ {\rm \mu m}$ wide Hall bars of single sided modulation doped GaAs/AlGaAs heterostructures containing  an 80, 60 or 50 nm wide QW. The spacer
thickness is 66 nm for the 80 and 60 nm QW samples and 82 nm for the thinnest QW. The samples are overdoped either using a short period superlattice doping (60 and 80 nm QW)~\cite{Umansky20091658} or conventional DX center doping (50 nm QW). An in-situ grown doped QW, located 800 nm below the 2DES, serves as a  backgate (BG) to tune the electron density. Resistance measurements were performed with standard lock-in technique in a dilution refrigerator with a base temperature below 20 mK.

\begin{figure*}[t]%
\includegraphics[scale=1]{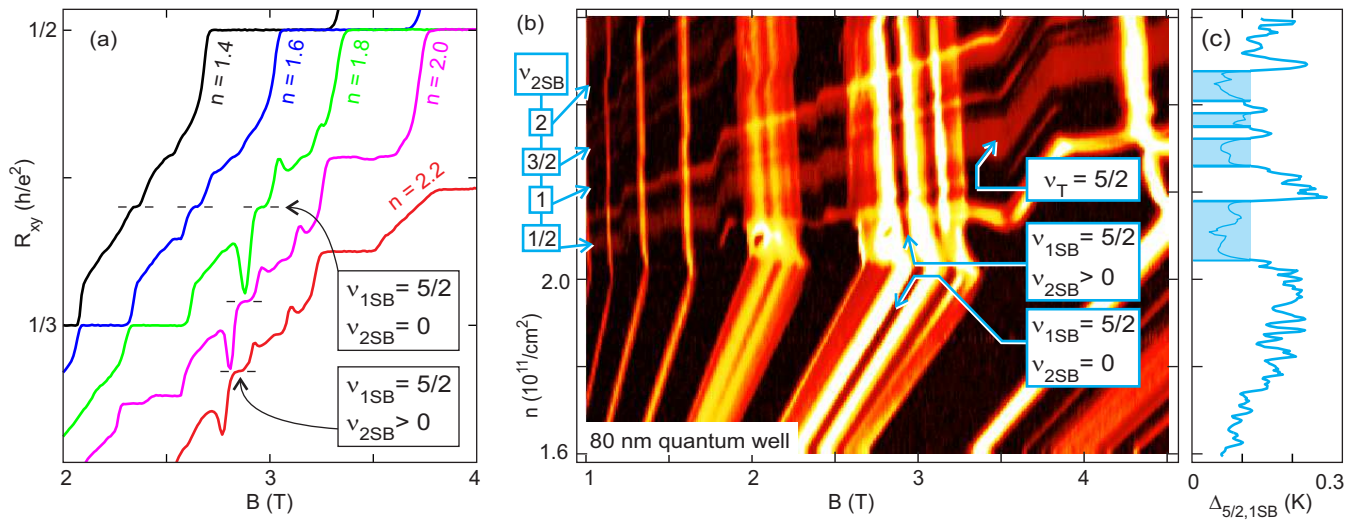}%
\caption{(Color) Transport data of a 2DES residing in a 80 nm wide quantum well. An in-situ grown backgate allows a variation of the electron density. (a) Hall resistance for selected densities (given in $10^{11}/\rm cm^2$). (b) Longitudinal resistance (see Fig.~\ref{60and50nm}(b) for a colorbar). Upon occupation of the second subband its electrons show integer and fractional quantum Hall states (examples indicated on the left). Meanwhile, the 5/2 state of the lower subband continues to exist. Its quantization is apparent from a plateau in $R_{xy}$, indicated in panel (a), and from vanishing $R_{xx}$ as seen in panel (b). The 5/2 state of the total density looses its quantization. (c) Activation energy of the lower subband 5/2 state. Regions where it is influenced by second subband features are shaded.}%
\label{80nm}%
\end{figure*}

Figure~\ref{80nm}(b) shows the longitudinal resistance in the density versus magnetic field plane measured on the 80 nm quantum well sample. Panel (a) shows the Hall resistance for selected densities. At lower densities electrons occupy only the 1SB and a 5/2 quantized Hall state is observed. At higher $n$ when also the 2SB is populated the 5/2 state for the {\em total} electron density looses its quantization. The longitudinal resistance no longer vanishes and the Hall plateau disappears in accordance with previous observations~\cite{PhysRevLett.105.246805, liu_robustness_52}. The 5/2 state of the 1SB, however, persists as an incompressible quantum Hall state over a wide range of fillings of the 2SB: the longitudinal resistance still vanishes and the Hall resistance still shows a plateau at filling factors ($\nu_{\rm 1SB} = 5/2$, $\nu_{\rm 2SB} > 0$). Yet, the plateau is found at progressively lower $R_{xy}$ as is expected when the total density increases. Meanwhile the 2SB electrons go through a series of integer and fractional quantum Hall states as indicated in panel (b). The overall structure shows that the fixed filling factors of the 1SB and 2SB correspond to a zigzag course. This can be understood as a consequence of intersubband charge transfer as addressed below and in~\cite{supplements}.

Hence, we have realized a compound system in which a 2DES condensing in a quantized 5/2 state coexists with a second 2DES whose density we can vary (in our current samples) between $0 < \nu_{\rm 2SB} \lesssim 1$. Here we investigate the transport properties of this system.
Even though the 5/2 state is known to be very fragile, we find it surprisingly undisturbed by the presence of the partially populated 2SB. Figure~\ref{80nm}(c) displays the activation energy (obtained from temperature dependent measurements) along the line of constant filling factor $\nu_{\rm 1SB} = 5/2$ as a function of the total density. In the $\nu_{\rm 2SB} = 0$ regime its absolute value and density dependence are comparable to our previous findings on a 30 nm QW~\cite{PhysRevB.81.035316}. When the 2SB is populated the activation energy $\Delta_{\nu_{\rm 1SB} =\rm 5/2}$ can only be determined where the electrons in the 2SB condense into an incompressible state and thus do not contribute to the longitudinal resistance (other regions are shaded in blue). There, $\Delta_{\nu_{\rm 1SB} =\rm 5/2}$ remains independent of the 2SB filling within the experimental uncertainty. This underlines that the only weakly changing density of the 1SB determines this state, while the influence from the 2SB is small.
We note that intersubband scattering may affect this state. However, in our samples the mobilities drop only by about one third upon population of the 2SB, indicating that intersubband scattering is not such a large effect.

\begin{figure}[t]%
\includegraphics[scale=0.98]{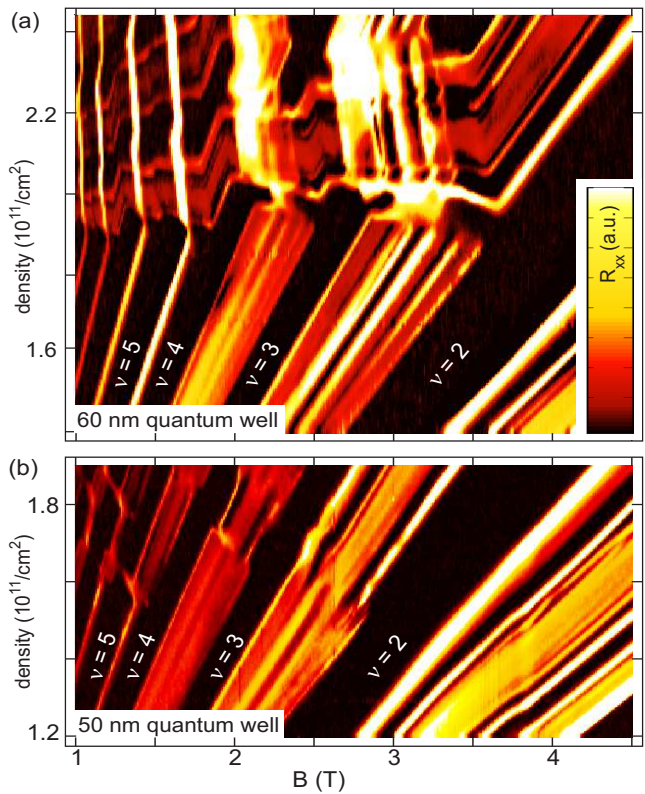}%
\caption{(Color) Longitudinal resistance. (a) In the 60 nm sample, when two subbands are populated their quantum Hall features coexist. In particular, a quantized 5/2 state of the lower subband is observed even when the second subband is populated. (b) In the 50 nm sample the quantum Hall features follow the total density.}%
\label{60and50nm}%
\end{figure}

Our observation of  a 5/2 state in the 1SB at a non integer filling of the 2SB implies that the topmost occupied LLs of both SBs are pinned at the same energy. Otherwise, it would be energetically favorable to transfer electrons from one SB to the other, filling up the lower of the partially populated levels (see~\cite{supplements} for details). But such intersubband charge transfer quickly aligns the topmost partially filled SB levels due to the additional capacitive energy involved. Charge transfer then stops until one of the Landau levels is completely filled. This energy depends strongly on the QW width. We address this theoretically below. We first look at the QW width dependence in experiment.
We observe the ($\nu_{\rm 1SB} = 5/2$, $\nu_{\rm 2SB} > 0$) state also in a 60 nm QW as illustrated in Fig.~\ref{60and50nm}(a). However, the behavior in a 50 nm QW (panel b) is different. In contrast to the wider QWs, no quantum Hall features associated with only the 1SB or 2SB show up once the 2SB is populated. Rather, nearly all features follow the {\em total} filling factor. The 5/2 quantized Hall state disappears once the 2SB gets populated. This is identical to what has been reported previously on QWs of up to 57 nm width~\cite{liu_thirds, PhysRevLett.87.196801}. We conclude that the QW width plays a crucial role in observing the incompressible 5/2 state when two SBs are occupied.

{\em Intersubband charge transfer - } As mentioned above, a partial filling of the topmost Landau levels of both SBs requires the alignment of these LLs. This is possible because these levels can be shifted with respect to each other when charges are transferred between the two subbands ~\cite{chargetransfer}. The individual SB densities (and wavefunctions) can be obtained by carrying out self-consistent calculations for each point in the density versus magnetic field plane, taking into account the $B-$dependent density of states (DOS)~\cite{PhysRevB.39.10232}. Here, we choose a different approach. We perform self-consistent calculations only for $B=0$ and treat the magnetic field induced effects perturbatively. This gives some insight into the importance of the QW width. The magnetic field induced changes in the DOS lead to oscillations in the subband splitting $\Delta_{\rm SB}$ (the difference in the lowest energy states of both SBs) as recently measured by photoluminescence~\cite{PhysRevB.80.241310}.

\begin{figure}%
\includegraphics[scale=1]{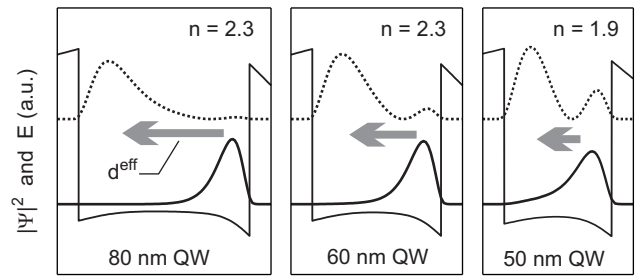}%
\caption{Probability densities ($|\Psi|^2$) of the first (solid) and second (dashed) subband. For the chosen total density (given in $10^{11}/\rm cm^2$) the second subband is populated. The effective spatial separation of the two subbands $d^{\rm eff}$ determines the capacitive energy needed for transferring electrons from one subband to the other. The quantum well is sketched. }%
\label{QWs}%
\end{figure}

For $B=0$ we self-consistently solve the Schr\"odinger and Poisson equation. Details can be found in~\cite{supplements}. Here, Fig.~\ref{QWs} shows the shape of the wavefunctions of the first and second SB. A representative density was chosen where the 2SB is populated. We observe that in the wider QWs the subband charge distributions are well separated while in the narrowest QW they are more centered and the spatial overlap is larger.

\begin{figure}%
\includegraphics[scale=1]{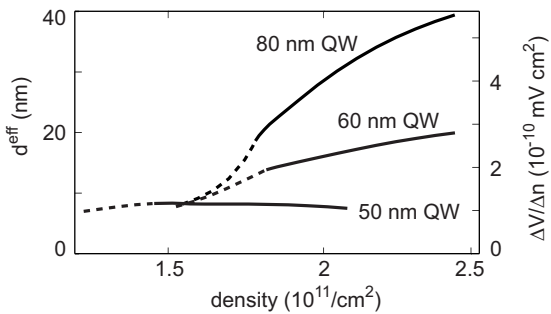}%
\caption{Effective spatial displacement when charge is transferred between two subbands for our samples. The right axes displays the shift between the subband energies per unit of areal density of transferred charge. In the dashed regions the second subband is not yet populated.}%
\label{deff}%
\end{figure}

A perpendicular magnetic field discretizes the DOS of both subbands into ladders of Landau levels. To minimize the total energy, charges may be redistributed between the two SBs. In analogy to charging a parallel plate capacitor, this leads to an energy shift between the SBs so that the topmost populated LLs of both SBs get aligned.  The energy shift between the two SBs when transferring a charge density  $e\Delta n$ from one
SB to the other can be described in terms of an effective capacitance per unit area $c^{\rm eff}$:
\begin{equation}
e\Delta V = \frac{e^2\Delta n}{c^{\rm eff}} = \frac{d^{\rm eff}}{\varepsilon \varepsilon_{0}}e^2\Delta n ,
\label{eq:energy_shift}
\end{equation}
where $\varepsilon \varepsilon_0$ is the dielectric constant of GaAs and $d^{\rm eff}$ is an effective distance. If the SB probability densities ($|\Psi_1(z)|^2$ and $|\Psi_2(z)|^2$ with $\Psi_{1,2}$ the wavefunctions of the 1SB and 2SB) are well separated the distance of their maxima can be used as an approximate value for $d^{\rm eff}$~\cite{PhysRevLett.79.2722}.
Here we calculate it for partially overlapping probability densities. The transfer of $e\Delta n$ between the two SBs creates a charge density $\rho(z)= e\Delta n (|\Psi_1(z)|^2 - |\Psi_2(z)|^2)$ inside the QW. The resulting Hartree energy per unit area is
$E_H = {1 \over 2} \int dz dz^\prime \rho(z) v(z-z^\prime) \rho(z^\prime)$, with  $v(z)=-|z|/(2 \epsilon \epsilon_0)$
denoting the Coulomb interaction between two homogeneously charged parallel planes.
The energy shift between the SBs is then $e\Delta V = \partial E_H/\partial (\Delta n)$, such that, with $f(z) = |\Psi_{1}(z)|^2 - |\Psi_{2}(z)|^2$, the effective distance can be written as
\begin{equation}
d^{\rm eff} = -  {1 \over 2} \int{dz \, dz' \,  f(z)   |z-z'| f(z^\prime)}.
\label{eq:deff}
\end{equation}
In Fig.~\ref{deff} we display results for our specific heterostructures and also convert $d^{\rm eff}$ into the energy shift per unit of transferred areal charge $\Delta V / \Delta n$. In the wider QW samples $d^{\rm eff}$ increases with the total density reflecting a progressively larger separation between the charges occupying the 1SB and 2SB. The 50 nm QW shows a qualitatively different behavior, namely $d^{\rm eff}$ is approximately constant and remains small when the 2SB starts to be populated.

From Fig.~4, we conclude that in general for narrower QWs more charges need to be transferred between the two SBs in order to align their topmost partially filled LLs. At the same time the total charge that can be transferred is limited by the degeneracy of the LLs and charge transfer will stop when either one of the SBs reaches the nearest integer filling factor. All in all, the parameter range in which LLs of both SBs are simultaneously only partially filled shrinks with decreasing QW width. Indeed, these regions are largest in our 80 nm QW and smaller in the 60 nm QW. Quantitative calculations for the 60 nm QW are described in~\cite{supplements}. The results agree well with the experiments, even though the influence of the intersubband charge transfer on the wavefunctions was not taken into account self-consistently. Also interaction effects were neglected. They cause, for example, a negative compressibility at low densities~\cite{PhysRevB.55.9294}, which can be seen in Fig. 1(b) at the onset of 2SB population as a kink in the slope of the 1SB features.

{\em Discussion - } We demonstrated that in sufficiently wide QWs the transfer of a small amount of charge from one SB to the other aligns their topmost partially filled LLs. The 2D electron systems  associated with each subband can condense in a fractional quantum Hall state while the other also
undergoes quantum Hall transitions. In particular, we observed an incompressible 5/2 state in the 1SB while the 2SB is being occupied. This additional 2DES of variable density is located in close proximity to the electrons forming a quantized 5/2 state,which is only weakly affected whenever the 2SB electrons condense into a FQHS. Future studies may focus on samples with an additional topgate. This would allow to tune in-situ the spatial separation of the SBs and thereby their Coulomb interaction and mutual screening properties.

Finally, we address the absence of the $\nu_{\rm 1SB} = 5/2$ fractional quantum Hall state in the 50 nm QW sample when the 2SB becomes occupied. Our calculations indicate that the capacitive energy for intersubband charge transfer is smaller than in the wider QWs, but not negligible, and regions of coexisting fractional SB filling factors should be observed. We conjecture that their absence is related to a larger overlap between the charge or probability densities $|\psi_{\rm 1SB}|^2$ and $|\psi_{\rm 1SB}|^2$ of the two subbands. This overlap enhances the probability for intersubband scattering. It promotes level mixing
which can result in an avoided level crossing so that an energy gap prevents the coexistence of a partially filled Landau level in both SBs. This would explain the observation that QH features are determined by the total density even when the 2SB is populated. An in-plane magnetic field has a similar effect~\cite{PhysRevB.78.233305}: It mixes the otherwise orthogonal 1SB and 2SB wavefunctions, provided that the probability densities associated with these wavefunctions overlap in space. Indeed, we observe in tilted field experiments (detailed results will be published elsewhere) that in the 80 nm QW a co-existence of partially filled LLs in both SBs is maintained even up to a tilt angle of $70^\circ$. In a 60 nm QW already at $10^\circ$ tilt angle, all single subband quantum Hall states have vanished once the 2SB becomes occupied. Hence, these tilted field measurements support the conjecture that a larger spatial overlap between the SB probability densities is responsible for the lack of single subband QH features once the 2SB becomes occupied in narrower QWs such as our 50 nm QW sample.

We acknowledge helpful discussions with S.~Dorozhkin and M.~Shayegan~\cite{liu_7_2} as well as financial support from GIF, DIP, ERC (FP7, grant 227716) and BMBF.

\end{document}